\documentclass[prl, twocolumn,amsmath,amssymb, superscriptaddress, nofootinbib,longbibliography]{revtex4-1}
\usepackage{graphicx}
\usepackage{natbib,hyperref,ulem,soul}
\usepackage{amsmath,epsfig,color,amssymb}
\usepackage{bibunits}

\definecolor{new}{rgb}{.08,.05,.8}

\defaultbibliography{NSFCareerfive}
\defaultbibliographystyle{apsrev4-1}

\begin{document}
\begin{bibunit}
	
\author{Areg Ghazaryan}
\affiliation{IST Austria (Institute of Science and Technology Austria), Am Campus 1, 3400 Klosterneuburg, Austria}

\author{P. L. S. Lopes}
\affiliation{Stewart Blusson Quantum Matter Institute, University of British Columbia,
	Vancouver, British Columbia, Canada V6T 1Z4}

\author{Pavan Hosur}
\affiliation{Department of Physics, University of Houston, Houston, TX 77204, USA}

\author{Matthew J. Gilbert}
\affiliation{
	Micro and Nanotechnology Laboratory, University of Illinois at Urbana-Champaign, Urbana, Illinois 61801, USA
}
\affiliation{
	Department of Electrical and Computer Engineering, University of Illinois at Urbana-Champaign, Urbana, Illinois 61801, USA
}
\affiliation{Department of Electrical Engineering, Stanford University, Stanford, California 94305, USA}
\date{\today}

\author{Pouyan Ghaemi}
\affiliation{Physics Department, City College of the City University of New York, New York, NY 10031, USA}
\affiliation{Graduate Center of the City University of New York, NY 10031, USA}

\title{Effect of Zeeman coupling on the Majorana vortex modes in iron-based topological superconductors}

\begin{abstract} 
In the superconducting regime of FeTe$_{(1-x)}$Se$_x$, there exist two types of vortices which are distinct by the presence or absence of zero energy states in their core. To understand their origin, we examine the interplay of Zeeman coupling and superconducting pairings in three-dimensional metals with band inversion. Weak Zeeman fields are found to suppress the intra-orbital spin-singlet pairing, known to localize the states at the ends of the vortices on the surface. On the other hand, an orbital-triplet pairing is shown to be stable against Zeeman interactions, but leads to delocalized zero-energy Majorana modes which extend through the vortex. In contrast, the finite-energy vortex modes remain localized at the vortex ends even when the pairing is of orbital-triplet form. Phenomenologically, this manifests as an observed disappearance of zero-bias peaks within the cores of topological vortices upon increase of the applied magnetic field. The presence of magnetic impurities in FeTe$_{(1-x)}$Se$_x$, which are attracted to the vortices, would lead to such Zeeman-induced delocalization of Majorana modes in a fraction of vortices that capture a large enough number of magnetic impurities. Our results provide an explanation to the dichotomy between topological and non-topological vortices recently observed in FeTe$_{(1-x)}$Se$_x$.
\end{abstract}
	
\maketitle

\textit{Introduction:} To date, one of the major impediments in the search for Majorana fermions (MFs) is that a requisite topological superconductivity, either intrinsic \cite{revtsc,Sato_2017} or induced in a host material via a proximity coupling to a standard $s$-wave superconductor \cite{PhysRevLett.100.096407,PhysRevLett.105.077001,Mourik25052012,Nadj-Perge602}. Of the available materials that possess topology, superconductivity and magnetism, iron-based superconductors are of recent interest \cite{revpnic,RevModPhys.87.855,PhysRevLett.121.076401,PhysRevB.100.014512,PhysRevB.88.134510,PhysRevB.90.024517,PhysRevB.98.144517,PhysRevLett.122.207001,dass}. In particular, the iron-based superconductor FeTe $_{0.55}$Se$_{0.45}$ (FTS) has recently been shown to have strong spin-orbit interactions and band inversion that result in a helical, topologically-protected, Dirac cone on the surface \cite{Hanaguri,PhysRevB.92.115119,Zhang182,Gangxu2016}. The phenomenology of vortices, proliferated in the presence of magnetic fields, is also noteworthy in FTS \cite{Wang333,tamagai,Lingyuan,hongjungao2019}. The  low charge density in the superconducting phase of this system is experimentally advantageous as it results in large Caroli-de Gennes-Matricon (CDM) vortex mode gaps \cite{CAROLI1964307} which facilitates the spectral detection of zero-energy vortex modes via scanning tunneling microscopy (STM). Intriguingly, vortices in FTS show two distinct types of behavior: topological, carrying zero energy states consistent with the presence of MF, and trivial vortices that lack the zero energy state but carry finite energy CDMs.
	
While the evidence of MFs in FTS has been observed, a comprehensive understanding of the salient physics of the vortex composition is lacking. More precisely, the energy spectra of the vortices follow different hierarchies relative to the CDM vortex gap, $\delta=\frac{\Delta^2}{\mu}$, where $\Delta$ is the bulk superconducting gap and $\mu$ is the chemical potential. The trivial vortex energy spectrum scales as $(n+1/2)\delta$, (with $n\in\mathbb Z$) which does not include the zero mode, whereas the topological vortices follow $n\delta$. Additionally, the percentage of vortices with zero-energy modes decreases as the perpendicular magnetic field increases, despite the fact that the inter-vortex distances are generically larger than the superconducting coherence length \cite{Wang333,tamagai,chinkai}. It was also shown that the distribution of vortices with and without zero mode has no correlation with the charge disorder on the surface of the material \cite{tamagai}. These features, on aggregate, suggest that the properties that distinguish the two classes of vortices stem from the bulk properties of individual vortex rather than those of the surface states. In particular the effects of magnetic field, beyond generation of vortices, might crucially affect the properties of superconducting state and the vortices. 
	
Motivated by experimental observations, we examine the effects of Zeeman coupling on the vortex modes in FTS. As the topological properties of FTS are driven by band inversion, we eschew more complex band models and utilize a simple model of a doped 3D time-reversal symmetric (TRS) topological insulator (TI) which has been used as an appropriate toy model to investigate the properties of vortices in FTS \cite{mvx,Zhang2019}.
An essential property of the electronic band structure in TRS TI is the presence of degenerate Fermi surfaces. Due to the strong spin-orbit coupling, Zeeman field splits the the degenerate Fermi surface into two helical Fermi surfaces with opposite helicity (see supplementary material). In this letter, we show that the split Fermi surfaces prefer an orbital-triplet superconducting pairing which delocalizes the MF modes at the ends of the vortices on the surface. To make direct connection with the dichotomy of vortices in FTS we note that the Zeeman field may result from the magnetic impurities along the vortex core \cite{PhysRevX.9.011033}. Interestingly, such magnetic Fe impurities are known to exists in FTS \cite{PhysRevLett.108.107002,Yin2015} and they are attracted to the vortices \cite{PhysRevX.9.011033}. In addition, increase of magnetic field naturally leads to enhancement of Zeeman coupling which further destabilize the topological vortices, as is experimentally observed. We should also note that neutron scattering measurements have shown the evidence of ferromagnetic clusters of Fe atoms in FTS \cite{PhysRevLett.108.107002}. Observation of the clusters of vortices, with and without zero energy vortex modes, would further support our theory.
	
\textit{Model Hamiltonian:} 
The two-orbital model of a 3D TRS TI is represented by the tight-binding Hamiltonian $\mathcal{H}_{\mathbf{k}}=\sum_{\mathbf{k}}\psi_{\mathbf{k}}^{\dagger}\left[\tau_x\textbf{d}_\textbf{k}.\boldsymbol{\sigma}+m_\textbf{k}\tau_z-\mu\right]\psi_{\mathbf{k}}$ where Pauli matrices $\sigma_i$ and $\tau_i$ act on spin and orbital space respectively, $d_{k_i}=2t \sin(k_i)$, $m_\textbf{k}=M+m_0\sum_i \cos(k_i)$ and $M$, $m_0$, $t$ are parameters of the model and $\mu$ is the chemical potential. By varying the parameters, this model Hamiltonian could represent both strong and week TRS TIs \cite{PhysRevB.81.045120}. The tight-binding model is used for the numerical calculations while our analytical results are based on the low-energy effective model
\begin{equation}\label{eq:hTI}\mathcal{H}=\hbar v_F\tau_x\pmb{\sigma}\cdot\mathbf{k}+m_k\tau_z, 
\end{equation} where $m_k=m+\epsilon k^2$ is the effective mass term and $\textbf{k}$ is the momentum relative to the center of the Brillouin zone. The trivial (topological) insulator corresponds to $m\epsilon>0$ ($<0$). Without loss of generality, we take $\hbar v_F = 1$.
	
With the Hamiltonian defined, we begin our analysis in the metallic phase, when $\mu>|m|$. The model displays two degenerate Fermi surfaces that split by a Zeeman field $\Delta_Z$. Since the bulk band structure gap is large compared with superconducting gap, we use the effective Hamiltonian resulting from projecting the Hamiltonian in Eq. (\ref{eq:hTI}) into the states at the two Fermi surfaces:
\begin{equation}\label{kh}
\mathcal{H}=\int d^{3}\mathbf{k}f_{\mathbf{k}}^{\dagger}\left[\left(E_\textbf{k}-\mu\right)\nu_0+\mathbf{d}_{\mathbf{k}}\cdot\boldsymbol{\nu}\right]f_{\mathbf{k}}.
\end{equation}
Here $\nu_i$ are the identity or Pauli matrices acting on the the space of two Fermi surfaces. The vector $\mathbf{d}_{\mathbf{k}}=\frac{\Delta_Z}{2}\left(\begin{array}{ccc}
-\frac{m_{\mathbf{k}}}{E_{\mathbf{k}}}\frac{k_{x}^{2}+k_{y}^{2}}{\left|\mathbf{k}\right|}, & 0, & \frac{k_{z}}{k}\end{array}\right)$ presents the Zeeman field and $f_\textbf{k}$ is the Fermion fields $\psi_\textbf{k}$ projected onto the Fermi surfaces (see supplementary materials). The two fermi spin-split fermi surfaces are identified by diagonalizing the projected Hamitonian (\ref{kh}) in the $\nu$ space.

Previous analysis \cite{PhysRevLett.105.097001} identified two types of superconductivity in doped TIs which are energetically favorable: intra-orbital spin singlet, $\int d^{3}\mathbf{r}\psi^{\dagger}\Delta i\tau_0\sigma_{y}\psi^{\dagger T}+H.c.$, and inter-orbital orbital-triplet spin-singlet, $\int d^{2}\mathbf{k}\psi_{\mathbf{k}}^{\dagger}i\tau_{x}\sigma_{y}\psi_{-\mathbf{k}}^{\dagger T}+H.c.$. Henceforth, we will refer to these superconducting pairings as intra-orbital singlet and inter-orbital triplet pairings, respectively.

Upon projection onto the Fermi surfaces the superconducting pairing potentials assume the following form,
\begin{equation}\label{p}\frac{\Delta_{\alpha}}{2} \int d^{3}\mathbf{k} f_{\mathbf{k}}^{\dagger}e^{-i\phi_\textbf{k}} \nu_\alpha f_{-\mathbf{k}}^{\dagger T},
\end{equation} 
where for the intra-orbital singlet pairing $\alpha=1$ and for inter-orbital triplet pairing $\alpha=0$.  
Examination of the superconducting pairing term in Eq.~(\ref{p}) in comparison with the kinetic Hamiltonian in Eq.~(\ref{kh}) shows that the intra-orbital singlet pairing, $\Delta_1$, corresponds to pairing electrons between different Zeeman split fermi surfaces. In contrast, the inter-orbital triplet pairing, $\Delta_0$, pairs electrons solely within each of the Zeeman split Fermi surfaces. Thus, the effect of Zeeman coupling is to break the TRS within the model and imbalance the two pairing potentials in the favor of $\Delta_0$ which couples the electrons solely within each Zeeman split Fermi surface \cite{youngseokkim2016}.

To examine the outlined effect of Zeeman coupling on the dominant form of superconducting pairing, we utilize a linear gap equation to determine the critical temperatures of the two superconducting pairings \cite{PhysRevLett.105.097001,nakosai2012,hashimoto2016,PhysRevB.90.045130}(see supplementary materials). 
The corresponding $U-V$ model, in which $U$ and $V$ are the intra and inter orbital interaction, leads to the equation :
\begin{equation}
\label{LinGap}
\mathrm{det}\left|\begin{array}{cc}
	U\bar{\chi}-1 & U\chi_{2} \\
	V\chi_{2} & V\chi_{1} -1
\end{array}\right|=1, \qquad
V\chi_{0}=1.
\end{equation}
Here $\bar{\chi}$, $\chi_1$ and $\chi_2$ are the superconducting susceptibilities characterizing intra-orbital spin-singlet pairing and $\chi_0$ describes the inter-orbital spin triplet pairing. $\bar{\chi}=-\int_{-w_D}^{w_D}\mathcal{D(\xi)}\tanh\left(\xi/2T\right)/2\xi d\xi$ is the standard $s$-wave susceptibility, $D(\xi)$ is the density of states and $w_D$ is the Debye frequency.

\begin{figure}
\includegraphics[width=7cm]{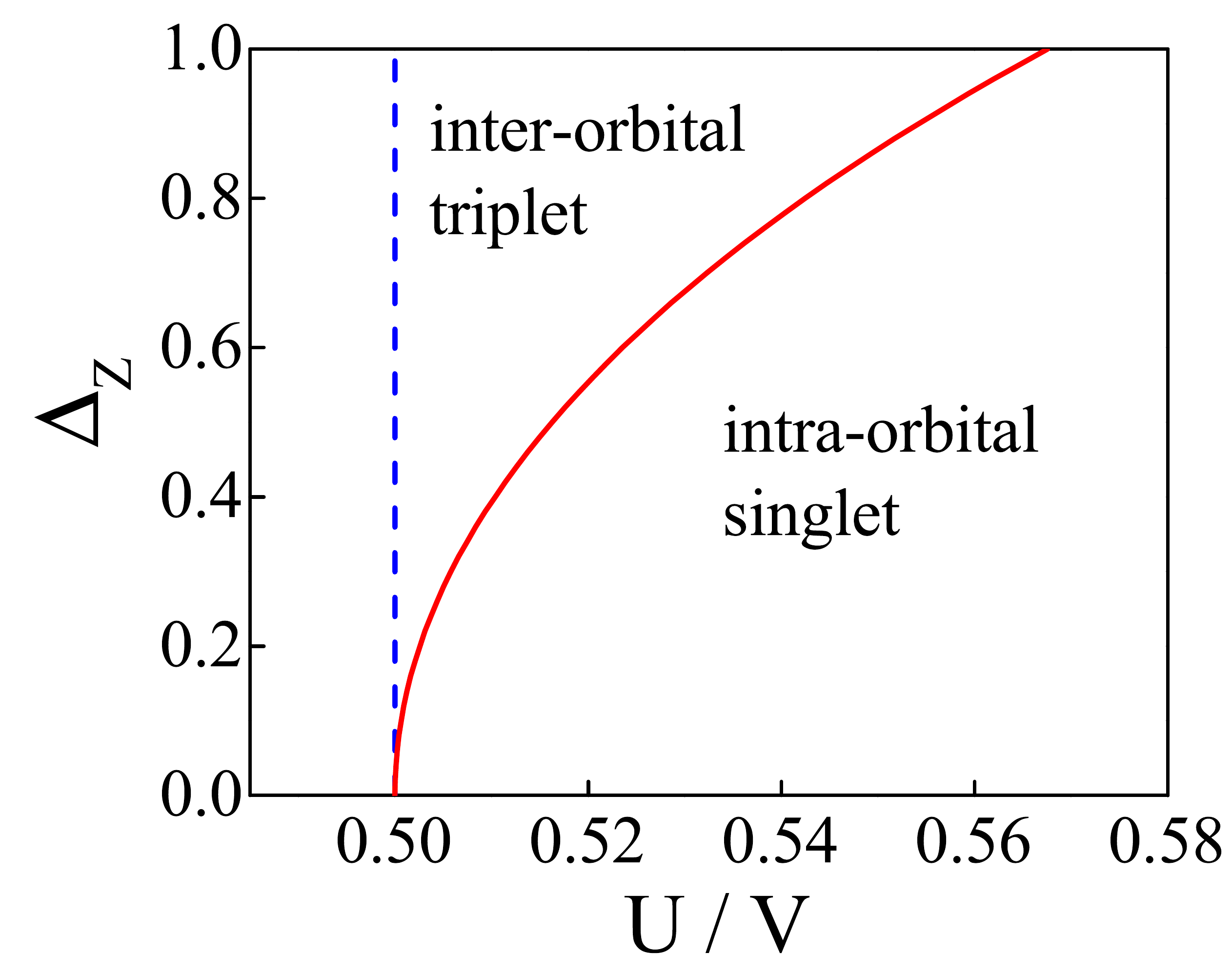}
\caption{\label{fig:PhaseSelfCons} Phase boundaries between regions with superconducting order parameters $\Delta_1$ or $\Delta_0$ as function of the ratio $U/V$ of interaction strengths of each channel and the field magnitude, $\Delta_Z$ (Red solid curve). The blue dashed line is guide for the eye of the case when $\Delta_Z=0$. The parameters of the Hamiltonian are $m=-0.5$ and $\epsilon=0.5$. Increased Zeeman coupling results in larger regions where inter-orbital triplet pairing is the ground state. \label{fig:phase_diagram}
}
\end{figure}

By numerically solving the $U-V$ Eq. (\ref{LinGap}), we obtain the critical temperature $T_c$ for each pairing channel. Fig.~\ref{fig:phase_diagram} shows the resulting phase boundaries that delineate the regions where either $\Delta_0$ or $\Delta_1$ correspond to higher critical temperature and so is the dominant form of pairing. In Fig.~\ref{fig:phase_diagram}, we plot the phase boundary as we vary $\Delta_Z$ and the ratio $U/V$. It is evident in Fig.~\ref{fig:phase_diagram} that the inclusion of the Zeeman effect results in the enhancement of the triplet ($\Delta_0$) pairing and the suppression of the singlet one ($\Delta_1$) at a given chemical potential. 
	
\textit{Vortex Modes:} Having established the phase diagram of the superconducting pairing, we proceed to study the internal structure of the vortices as we insert a $\pi$-flux-tube into the pairing potential: $\Delta_\alpha(r)=|\Delta_\alpha(r)|e^{i\theta}$~\cite{PhysRevLett.107.097001,chingkaichiu2011}. 

We fix $k_z=0$ as we are interested in points where the topological $\mathbb{Z}_2$ index changes and this only occurs at $k_z = 0$ or $\pi$. Since the vortex modes stem from states close to the Fermi energy, their wave functions can be expressed in terms of a superposition of the TI conduction band eigenstates in cylindrical coordinates $\chi_{l,k}^{\nu}$ (see supplementary material).

Here $l$s are angular momentum quantum numbers and $\nu=\pm$ correspond to the two energy bands of Hamiltonian (\ref{eq:hTI}) which are split by Zeeman coupling.

Given the rotational symmetry of the vortex profile, the vortex modes with different $l$s are not hybridized and the vortex Hamiltonian decouples into sectors associated with each $l$. Also for inter-orbital triplet pairing, the vortex does not mixes the $\nu$'s. On the other hand, the translation symmetry in the plane perpendicular to the vortex is broken and different radial momenta $k$, as well as Nambu particle-hole states are mixed by the vortex. The effective vortex Hamiltonian acting in the radial momentum and Nambu particle-hole spaces takes the form of a 1D Jackiw-Rebbi model \cite{PhysRevD.13.3398} (see supplementary material)
	
\begin{figure}
\includegraphics[width=8.5cm]{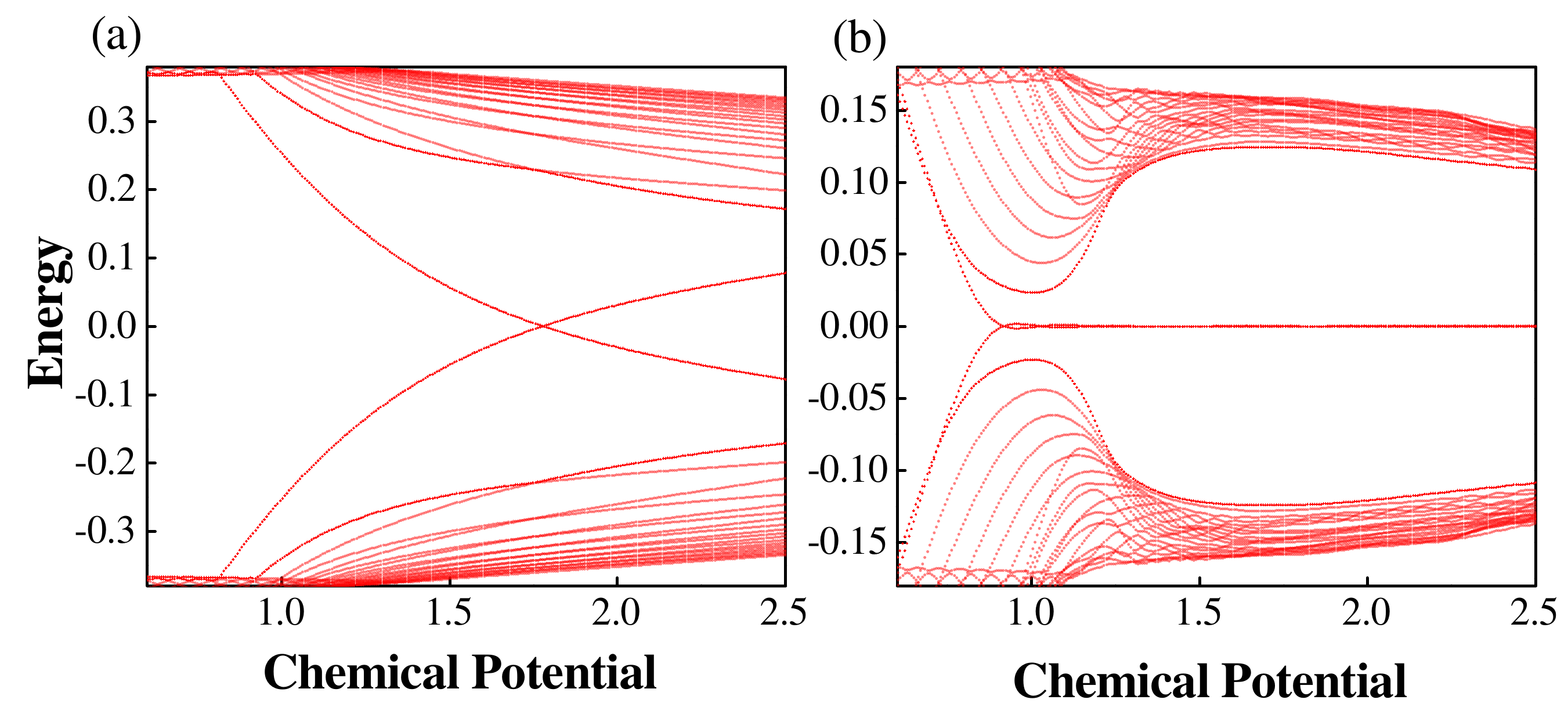}
\caption{Dependence of the vortex modes energies on the chemical potential $\mu$ obtained from the 3D lattice model with periodic boundary conditions along the $z$-direction for $k_z=0$. (a) Intra-orbital singlet pairing $\Delta_1=0.4$ for $\Delta_Z=0$. (b) Inter-orbital triplet pairing $\Delta_0=0.4$ and $\Delta_Z=0.2$.  The parameters for the model are: $M=4.5$, $m_0=-2.0$ and $t=1.0$ with the calculation performed on a $48\times48$ lattice in the $x-y$ plane. We observe a zero-energy mode state exists for all chemical potentials inside the conduction band when the pairing is inter-orbital triplet type.
} \label{fig:CdG_lattice}
\end{figure}

\begin{equation}
H_{l}^{\nu}=\Pi_{z}\frac{\Delta_{0}\lambda_{\nu}\left(k\right)}{\xi_0}+\Pi_{y}\frac{\Delta_{0}}{\xi_0}\left(i\partial_{k}\right)+\Pi_{x}E_{l,0}^{\nu}(k).\label{eq:JRHamilt}
\end{equation}
The $\Pi$ matrices act on a Nambu space, $\xi_0$ is the superconducting coherence length and $\Delta_{0}\lambda_{\nu}\left(k\right)/\xi_{0}=k+\frac{hm_{\mathbf{k}}}{2E^0_{\mathbf{k}}}\nu-\mu$. The Jackiw-Rebbi lowest-energy solutions are localized at the Fermi surface where the coefficient $\lambda_{\nu}\left(k_{F} ^\nu\right)$ changes sign. These states have the form $ e^{-\int_{k_{F}^{\nu}-k}^{k_{F}^{\nu}+k}dk'\lambda_{\nu}\left(k'\right)} \left(1,1\right)^T$ in Nambu particle-hole basis and has energy  $E_{l,0}^{\nu}(k_F)$. Notice that vortex-mode wave functions are exponentially localized around Fermi wavevector. Since the BdG Hamiltonian in equation (\ref{eq:JRHamilt}) is in the bases of TI conduction band states, the full wave function of the two vortex modes (each associated with one Zeeman split Fermi surface) takes the form \begin{equation}\label{vm}\Psi^\nu_V\left(l,r\right) \approx \chi^\nu_{l,k_F}(r)\left(1,1\right)_\Pi^T\end{equation}where $\left(1,1\right)_\Pi^T$ is the spinor in Nambu particle-hole space. Previously, a similar result was obtained for the intra-orbital spin-singlet case~\cite{PhysRevLett.107.097001,chingkaichiu2011}. In that case, solutions where again centered at the metallic phase Fermi surface, with corresponding energies $E^\nu_{l,1}=\frac{\Delta}{k_F \xi_1 }(2\pi l+\pi+\nu \phi_B)$. $\phi_B$ is a Berry-phase-like term which permits zero-modes whenever $\phi_B=\pi$. 
In contrast, for the inter-orbital triplet case, the energies of the vortex modes are $E_{l,0}^{\nu}=\frac{\Delta_0}{ k_{F}^{\nu}\xi_0}\left(2\pi l\right)$. They noticeably lack the Berry phase term present in the intra-orbital singlet case and the $l=0$ states always has zero energy. Therefore, in the inter-orbital triplet case, a zero-energy channel {\it always} exists {\it along} the vortex. Its presence delocalizes the zero-modes at the {\it end} of the vortex, on the sample surface. Delocalization of the zero mode at the ends of the vortex results in a suppression of the zero-bias signal in STM measurements.

In Fig.~\ref{fig:CdG_lattice} we verify the analytic results using a 3D lattice model with periodic boundary conditions along the z-direction.
Fig.~\ref{fig:CdG_lattice} (a) shows the spectrum of the vortex modes for the intra-orbital singlet pairing where the vortex gap closes solely when $\phi_B = \pi$. In contrast, Fig.~\ref{fig:CdG_lattice} (b) shows that for the inter-orbital triplet pairing, once the chemical potential is in the conduction band the system develops vortex zero modes which remain gapless for all chemical potentials. Thus, increasing the Zeeman coupling results in a shift of the Fermi surfaces that destabilize the intra-orbital singlet pairing in favor of the inter-orbital triplet pairing leading to the omnipresence of a zero mode in the vortex core.  
	
\begin{figure}
\includegraphics[width=8cm]{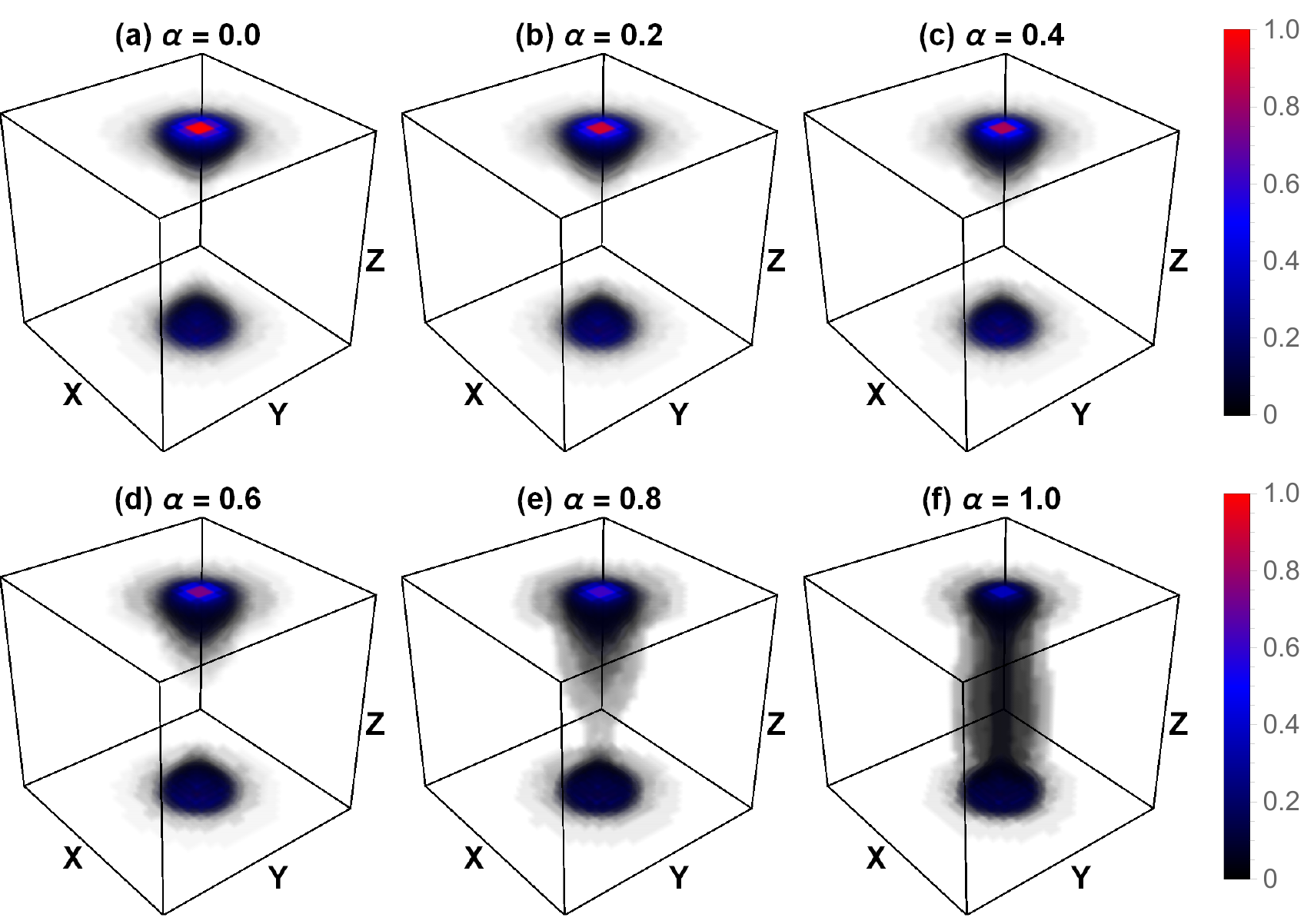}
\caption{\label{fig:3DWaveFunc}
Spatial profile of the lowest energy vortex mode in 3D slab geometry for different values of parameter $\alpha$, which controls the amplitude of the inter-orbital triplet pairing $\Delta_0=\alpha\Delta^0_0$, intra-orbital s-wave $\Delta_1=(1-\alpha)\Delta^0_1$ and Zeeman field $\Delta_Z=\alpha\Delta^0_Z$ around the vortex. 
The calculation is performed on a $24\times24\times24$. The parameters used are: $\mu=1.1$, $\Delta^0_0=0.4$, $\Delta^0_1=0.4$,  $\Delta^0_Z=0.2$, $M=4.5$, $m_0=-2.0$ and $t=1.0$.}
\end{figure}

In Fig.~\ref{fig:3DWaveFunc}, we examine the manifestation of the change in superconductive pairing by numerically inserting a vortex in the tight-binding lattice model. We set $\mu = 1.1 t$, where $t$ is hoping amplitude in the lattice model. For small Zeeman couping, the intra-orbital singlet channel is dominant and the vortex states, both topological and trivial, are localized at the ends of the vortex on the surface. As the Zeeman splitting is increased, we destabilize the intra-orbital singlet pairing channel in favor of the inter-orbital triplet pairing allowing MFs, localized at the ends of the vortex string, to penetrate into the bulk. At sufficiently high Zeeman coupling, the intra-orital singlet pairing is fully suppressed and MFs on the surface delocalize through the vortex modes and extend into the bulk of the superconductor. We may also understand this mechanism, starting from the case where the pairing is solely inter-orbital triplet type where the vortex hosts two zero modes extended through the bulk. Intra-orbital singlet pairing hybridizes these two modes and generate a gap for the vortex modes extending through the bulk and localizes the MFs at the ends of the vortex on the surface. Interestingly, as will be shown below, this mechanism would only delocalize the Majorana zero modes and could not do the same for finite energy vortex modes at the ends of the vortex. It is then well consistent with the experimental results which shows that finite energy vortex modes are intact in all vortices, whereas the MFs are absent in some of the vortices.

\textit{Effective 1D Vortex Model:} Using the wavefunctions from Eq.~\eqref{vm}, we now derive an effective Hamiltonian for the modes along the vortex line connecting the surfaces ~\cite{PhysRevB.92.064518}. To this end, we calculate the matrix elements of the $k_z$-dependent terms in  Hamiltonian, Eq.~\eqref{eq:hTI}, within the space of the two vortex zero modes presented in Eq.~(\ref{vm}). Without lose of generality we consider $m>0$ and $\epsilon<0$. Defining $\eta_i$ as Pauli matrices acting on the space of the two zero modes, the effective vortex Hamiltonian have the general form
\begin{align}
\label{eq:od}
H^l_{V} &=\frac{E^+_{l}+E^-_{l}}{2}\eta_0 \\
&+\left(\frac{E^+_{l}-E^-_{l}}{2}-\tilde{\epsilon}\partial_z^2\right)\eta_z+\tilde{v}_z\left(\partial_z\right) \eta_x \nonumber.
\end{align}
Here, $E^{\nu}_{l}$ are the energies of $l$-th vortex modes from Fermi surface corresponding to $\nu=\pm$. In Eq.~\eqref{eq:od}, both $\tilde{\epsilon}$ and $\tilde{v}_z$ are parameters of the $k_z$-dependent terms. The effective one dimensional Hamiltonian in Eq.~\eqref{eq:od} corresponds to the effective vortex Hamiltonian and supports localized states at the two ends if $\left(E^+_l-E^-_l\right)\tilde\epsilon<0$~\cite{RevModPhys.83.1057}.

\begin{figure}
\includegraphics[width=8.5cm]{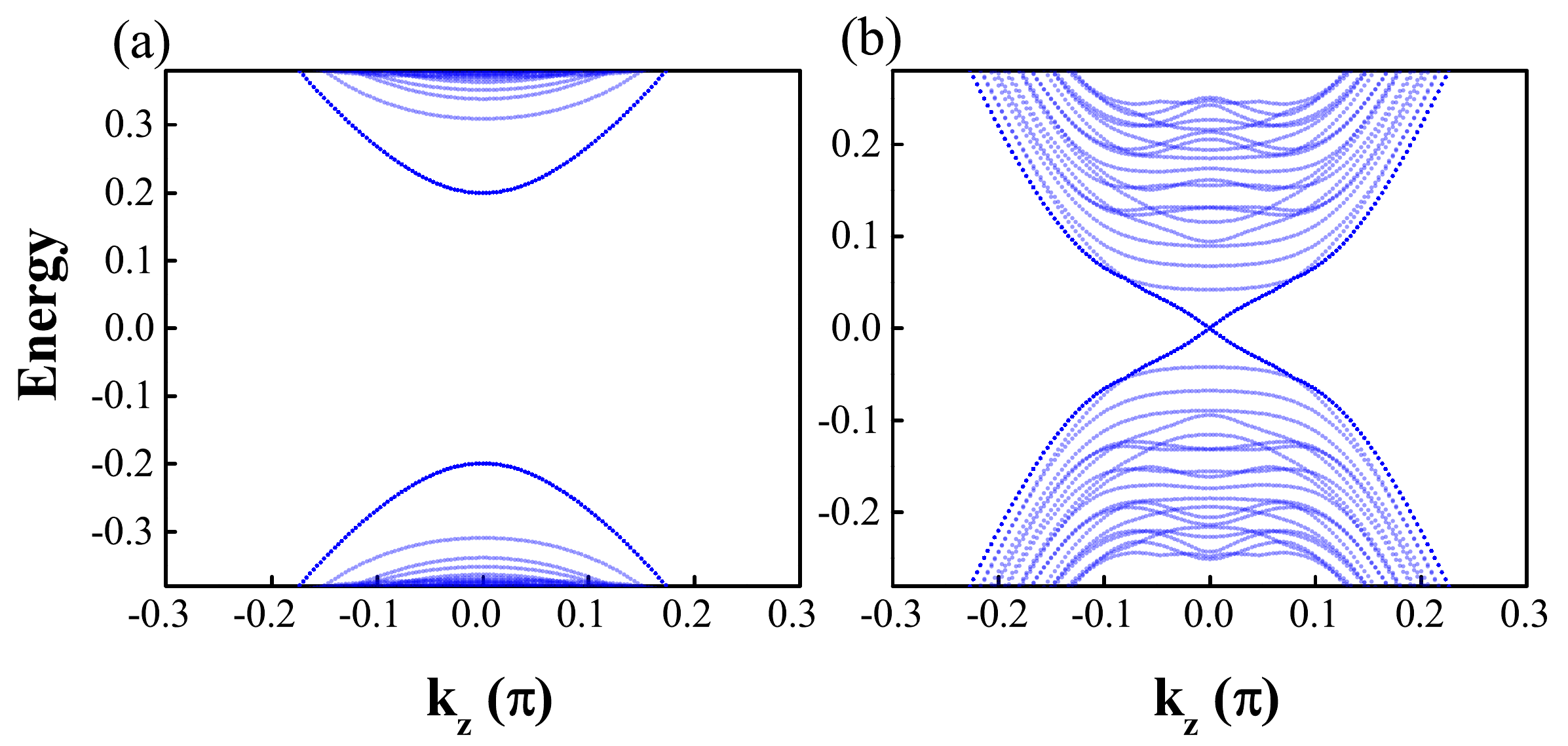}
\caption{\label{fig:EnDepKz} $k_z$ momentum dependence of the vortex modes for (a) intra-orbital $s$-wave pairing $\Delta_1=0.4$ and (b) inter-orbital triplet pairing $\Delta_0=0.4$ with $\Delta_Z=0.2$ and the chemical potential set to be $\mu=1.1$. Additional parameters of the model are the same as in Fig.~\ref{fig:CdG_lattice}.}
\end{figure}

From the definition of $\lambda_{\nu}\left(k_{F} ^\nu\right)$, we notice $k_F^+<k_F^-$ and consequently, $E_{l}^+-E_{l}^->0$. Therefore, the condition for the presence of localized states at the end of vortex with finite energy $\left(E^+_{l}-E^-_{l}\right)\tilde{\epsilon}<0$ is satisfied. The modes with $l\neq 0$ and finite energies of $\frac{E^+_{l}+E^-_{l}}{2}$ are localized at the ends of the vortex lines. In contrast, for the two zero energy modes ($l=0$) with $\left(E^+_{0}-E^-_{0}\right)\tilde{\epsilon}=0$. As a result, the MFs can not stay localized at the ends of the vortex when the bulk pairing is of inter-orbital triplet form. As shown in Fig.~\ref{fig:EnDepKz}(b), the effective vortex Hamiltonian for the zero energy states is linearly dispersing with momentum along the vortex and, due to their Nambu particle-hole character, protected against back-scattering which further supports the delocalization of zero modes upon formation of inter-orbital triplet pairing.

\textit{Conclusion:} We have examined the effect of Zeeman coupling on the structure of vortex modes in FTS. We find that the inter-orbital triplet paring and intra-orbital singlet pairing compete as a function of Zeeman coupling, resulting in dramatically different vortex structures. Intra-orbital singlet pairing leads to the presence of Majorana vortex modes localized on the surface at the end of vortex strings.  On the other hand, inter-orbital triplet paring supports localized finite-energy trivial vortex modes but destabilize the zero-energy Majorana modes at the end of vortex strings. This delocalization is through the formation of zero modes which extend along the vortex through the bulk. Such extended states have small coupling with STM probe and do not show strong signals, in comparison with the case where the zero modes are localized at the end of vortex, on the sample surface. Our results shed light in the existence of two types of vortices experimentally observed in FTS and support the topological nature of vortex modes in this system.
	
The main property of the band structure of FTS to facilitate our models is the presence of degenerate Fermi surfaces that are split into two helical Fermi surfaces by the Zeeman field. The size of the superconducting gap in FTS is of the order of $2.5 \,\mathrm{meV}$ for hole pocket and $4.2 \,\mathrm{meV}$ for electron pocket  \cite{PhysRevLett.108.107002,Yin2015,Zhang182}. The size of the Fe impurities dipole-moments in FTS is of the order of $5 \mu_B$ \cite{PhysRevLett.108.107002,PhysRevX.9.011033}. Given the average distance of the Fe impurity atoms, their associated Zeeman coupling is of the order of $\Delta_Z \approx 7.84 \,\mathrm{meV}$. It is then evident that the Zeeman coupling can affect the form of the superconducting paring. Our results then demonstrate that the nature of the vortices in FTS is inextricably linked to the effect of Zeeman coupling, which determines the form of superconducting pairing. It also indicates that suppression of Zeeman coupling, by reduction of magnetic impurities, stabilizes the vortex MFs in FTS.

\textit{Acknowledgments.} 
This research was supported under National Science Foundation Grants EFRI-1542863 and PSC-CUNY Award, jointly funded by The Professional Staff Congress and The City University of New York (PG). PLSL acknowledges support by the Canada First Research Excellence Fund. AG acknowledges support from the European Union’s Horizon 2020 research and innovation program under the Marie Skłodowska-Curie grant agreement No 754411. PH was supported by the Department of Physics and the College of Natural Sciences and Mathematics at the University of Houston. M.J.G. acknowledges financial support from the National Science Foundation (NSF) under Grant No. DMR-1720633, CAREER Award ECCS-1351871 and the Office of Naval Research (ONR) under grant N00014-17-1-3012.

\putbib
\end{bibunit}

\begin{bibunit}
	
\clearpage
\onecolumngrid
\setcounter{figure}{0}
\makeatletter
\renewcommand{\thefigure}{S\@arabic\c@figure}
\setcounter{equation}{0} \makeatletter
\renewcommand \theequation{S\@arabic\c@equation}
\renewcommand \thetable{S\@arabic\c@table}

\begin{center}{\Large \textbf{Effect of Zeeman coupling on the Majorana vortex modes in iron-based topological superconductors - Supplementary Material}}\end{center}

In  this supplementary material, we present (a) the derivation of the vortex spectrum in doped topological insulator with intra-orbital singlet and inter-orbital triplet superconductivity, (b) a derivation of effective vortex Hamiltonian which captures the the properties of localized stats at the ends of the vortex (c) the self-consistent calculations to determine the effect of Zeeman coupling on the type of superconducting state. 

\section{Vortex modes in doped topological insulators}

\subsection{Model Hamiltonian}

The effective Hamiltonian from Eq. (1) of the main text is $\mathcal{H}_{0}=\int d^{3}r\psi^{\dagger}\left(\mathbf{r}\right)\left(H-\mu\right)\psi\left(\mathbf{r}\right)$ 
where $\psi\left(\mathbf{r}\right)=\left(\psi_{\uparrow A},\psi_{\downarrow A},\psi_{\uparrow B},\psi_{\downarrow B}\right)^{T}$. Here $\psi_{\sigma J}$ is the destruction operator for electron in orbital $J$ with spin $\sigma$.
In momentum space, the Hamiltonian reads 
\begin{equation}
	\mathcal{H}_{0}=\int d^{3}\mathbf{k}/\left(2\pi\right)^{3} \psi_{\mathbf{k}}^{\dagger}\left(H_{\mathbf{k}}-\mu\right)\psi_{\mathbf{k}},
\end{equation}
where
\begin{equation}\label{mh}
	H_{\mathbf{k}}=\boldsymbol{\alpha}\cdot\mathbf{k}+m_{\mathbf{k}}\beta
\end{equation}
with $m_{\mathbf{k}}=m-\epsilon k^{2}$,  $\boldsymbol{\mathbf{\alpha}}=\tau_{x}\boldsymbol{\sigma}$ and
$\beta=\tau_{z}\sigma_{0}$. $\tau$ and $\sigma$ matrices act on
the $A,B$ orbitals and $\uparrow\downarrow$ spins, respectively.

The dispersions of the four bands follow $E_{\mathbf{k}}^{1}=E_{\mathbf{k}}^{2}=-E_{\mathbf{k}}^{3}=-E_{\mathbf{k}}^{4}\equiv E_{\mathbf{k}}=\sqrt{k^{2}+m_{\mathbf{k}}^{2}}$.
We expand the Fermion operators in an eigenbasis as
\begin{equation}
	\psi_{\mathbf{k}}=\sum_{a=1}^{4}\varphi_{\mathbf{k}}^{a}f_{\mathbf{k},a},
\end{equation}
where the eigenstates $\varphi_{\mathbf{k}}^{a}$'s are given by
\begin{equation}
	\varphi_{\mathbf{k}}^{a}=e^{i\phi_{\mathbf{k}}/2}e^{-i\sigma_{z}\phi_{\mathbf{k}}/2}e^{-i\sigma_{y}\theta_{\mathbf{k}}/2}e^{-i\tau_{y}\sigma_{z}\alpha_{\mathbf{k}}/2}\hat{e}_{a},
\end{equation}
with 
\begin{align}
	\tan\phi_{\mathbf{k}} & =\frac{k_{y}}{k_{x}}\nonumber \\
	\cos\theta_{\mathbf{k}} & =\frac{k_{z}}{k}\nonumber \\
	\cos\alpha_{\mathbf{k}} & =\frac{m_{\mathbf{k}}}{E_{\mathbf{k}}},
\end{align}
where $\hat{e}_{a}$ is a basis of $\mathbb{R}^{4}$ corresponding to the four bands. The phases
are chosen so that the wavefunctions are single valued upon a $2\pi$
evolution of the azimuthal momentum-angle variable. The angle-variables
transform according to $\mathbf{k}\to-\mathbf{k}$ as $\phi_{-\mathbf{k}}=\phi_{\mathbf{k}}-\pi$
and $\theta_{-\mathbf{k}}=\pi-\theta_{\mathbf{k}}$, while $\alpha_{-\mathbf{k}}=\alpha_{\mathbf{k}}$.
Notice $\alpha_{\mathbf{k}}$ depends only on the modulus of $\mathbf{k}$. Explicitly, the two upper-band wavefunctions are
\begin{align}
	\varphi_{\mathbf{k}}^{1} & =e^{i\phi_{\mathbf{k}}/2}e^{-i\sigma_{z}\phi_{\mathbf{k}}/2}e^{-i\sigma_{y}\theta_{\mathbf{k}}/2}e^{-i\tau_{y}\sigma_{z}\alpha_{\mathbf{k}}/2}\left(\begin{array}{c}
		1\\
		0\\
		0\\
		0
	\end{array}\right)\nonumber \\
	& =\left(\begin{array}{c}
		\cos\frac{\theta_{\mathbf{k}}}{2}\cos\frac{\alpha_{\mathbf{k}}}{2}\\
		e^{i\phi_{\mathbf{k}}}\sin\frac{\theta_{\mathbf{k}}}{2}\cos\frac{\alpha_{\mathbf{k}}}{2}\\
		\cos\frac{\theta_{\mathbf{k}}}{2}\sin\frac{\alpha_{\mathbf{k}}}{2}\\
		e^{i\phi_{\mathbf{k}}}\sin\frac{\theta_{\mathbf{k}}}{2}\sin\frac{\alpha_{\mathbf{k}}}{2}
	\end{array}\right)
\end{align}
and

\begin{align}
	\varphi_{\mathbf{k}}^{2} & =e^{i\phi_{\mathbf{k}}/2}e^{-i\sigma_{z}\phi_{\mathbf{k}}/2}e^{-i\sigma_{y}\theta_{\mathbf{k}}/2}e^{-i\tau_{y}\sigma_{z}\alpha_{\mathbf{k}}/2}\left(\begin{array}{c}
		0\\
		1\\
		0\\
		0
	\end{array}\right)\nonumber \\
	& =\left(\begin{array}{c}
		-\sin\frac{\theta_{\mathbf{k}}}{2}\cos\frac{\alpha_{\mathbf{k}}}{2}\\
		e^{i\phi_{\mathbf{k}}}\cos\frac{\theta_{\mathbf{k}}}{2}\cos\frac{\alpha_{\mathbf{k}}}{2}\\
		\sin\frac{\theta_{\mathbf{k}}}{2}\sin\frac{\alpha_{\mathbf{k}}}{2}\\
		-e^{i\phi_{\mathbf{k}}}\cos\frac{\theta_{\mathbf{k}}}{2}\sin\frac{\alpha_{\mathbf{k}}}{2}
	\end{array}\right).
\end{align}

We are going to assume a positive chemical potential, tuned inside
the upper two bands. This allows neglecting the lower bands with negative
energy as $\psi_{\mathbf{k}}=\sum_{a=1}^{4}\varphi_{\mathbf{k}}^{a}f_{\mathbf{k},a}\approx\sum_{a=1,2}\varphi_{\mathbf{k}}^{a}f_{\mathbf{k},a}$. This leads to

\begin{equation}
	\mathcal{H}_{0}\approx\sum_{a=1,2}\int_{\mathbf{k}}\left(E_{\mathbf{k}}-\mu\right)f_{\mathbf{k},a}^{\dagger}f_{\mathbf{k},a}.
\end{equation}where $f_{\mathbf{k}}=\left(f_{\mathbf{k},1},f_{\mathbf{k},2}\right)^{T}$
contains the two positive energy modes only. Notice that the conduction band consists of two degenerate bands.

\subsection{Zeeman coupling}

The Zeeman magnetization is incorporated through the operator $
\mathcal{H}_{Z}=\frac{\Delta_Z}{2}\int d^{3}r\psi^{\dagger}\left(\mathbf{r}\right)\sigma_{z}\psi\left(\mathbf{r}\right)$. Projecting in the two upper bands, we have to consider only the matrix
elements for $\hat{e}_{1}$ and $\hat{e}_{2}$. Introducing Pauli
matrices $\boldsymbol{\nu}$ to represent the two Fermi surfaces in the two degenerate conduction bands, the projected Zeeman term takes the form
\begin{equation}
	\mathcal{H}_{Z}\approx\int d^{3}\mathbf{k}f_{\mathbf{k}}^{\dagger}\left[\mathbf{b}_{\mathbf{k}}\cdot\boldsymbol{\nu}\right]f_{\mathbf{k}},
\end{equation}
where $\mathbf{b}_{\mathbf{k}}=\frac{\Delta_Z}{2}\left(\begin{array}{ccc}
-\cos\alpha_{\mathbf{k}}\sin\theta_{\mathbf{k}}, & 0, & \cos\theta_{\mathbf{k}}\end{array}\right)$. 

The signature of a vortex phase transition is the appearance of zero-energy vortex modes with zero momentum along the vortices. In what follows, we will thus consider
$k_{z}=0\Rightarrow\theta_{\mathbf{k}}=\pi/2$ that corresponds to $\mathbf{b}_{\mathbf{k}}=\frac{\Delta_Z}{2}\left(\begin{array}{ccc}
-\cos\alpha_{\mathbf{k}}, & 0, & 0\end{array}\right)\equiv -b^x_{\mathbf{k}}\hat{x}$. The Hamiltonian, including the projected Zeeman coupling then reads

\begin{equation}\label{zz}
	\left[\mathcal{H}_{0}+\mathcal{H}_{Z}\right]_{k_{z}=0}\approx\int d^{2}\mathbf{k}f_{\mathbf{k}}^{\dagger}\left[\left(E_{\mathbf{k}}-\mu\right)\nu_{0}-b^x_{\mathbf{k}}\nu_{x}\right]f_{\mathbf{k}}.
\end{equation}
It is evident that the Zeeman field mixes the modes of the two upper bands and can be diagonalized by the unitary transformation $d_{\mathbf{k}}=\left(d^+_{\mathbf{k}},d^-_{\mathbf{k}}\right)^T=e^{i\nu_{y}\pi/4}f_{\mathbf{k}}$.  Hamiltonian \eqref{zz} then becomes
\begin{equation}
	\left[\mathcal{H}_{0}+\mathcal{H}_{Z}\right]_{k_{z}=0}\approx\int d^{2}\mathbf{k}d_{\mathbf{k}}^{\dagger}\left[\left(E_{\mathbf{k}}-\mu\right)\nu_{0}-b^x_{\mathbf{k}}\nu_{z}\right]d_{\mathbf{k}}.
\end{equation}

\subsection{Effect of Zeeman coupling on intra-orbital singlet pairing}

The uniform intra-orbital s-wave pairing has the form:

\begin{align}
	\mathcal{H}_{SC,1} & =\int d^{3}\mathbf{r}\left(\psi_{\uparrow A}^{\dagger}\Delta_{1}\psi_{\downarrow A}^{\dagger}+\psi_{\uparrow B}^{\dagger}\Delta_{1}\psi_{\downarrow B}^{\dagger}\right)+H.c.\nonumber \\
	& =\frac{1}{2}\int d^{3}\mathbf{r}\psi^{\dagger}\Delta_{1}i\sigma_{y}\psi^{\dagger T}+H.c..
\end{align}

In momentum space and projecting in the upper Fermi surfaces,
\begin{align}
	\mathcal{H}_{SC,1} & =\frac{\Delta_{1}}{2}\int d^{2}\mathbf{k}\psi_{\mathbf{k}}^{\dagger}i\sigma_{y}\psi_{-\mathbf{k}}^{\dagger T}+H.c.\nonumber \\
	& \approx\frac{\Delta_{1}}{2}\sum_{a,b=1,2}\int d^{2}\mathbf{k}f_{\mathbf{k},a}^{\dagger}\left(\varphi_{\mathbf{k}}^{a\dagger}i\sigma_{y}\varphi_{-\mathbf{k}}^{b*}\right)f_{-\mathbf{k},b}^{\dagger}+H.c..
\end{align}

For $a,b=1,2$ and with angle-variable transformations $\mathbf{k}\to-\mathbf{k}$ results in

\begin{equation}
	\mathcal{H}_{SC,1}=-\frac{\Delta_{1}}{2}\int d^{2}\mathbf{k}\left(f_{\mathbf{k}}^{\dagger}e^{-i\phi_{\mathbf{k}}}\nu_{z}f_{-\mathbf{k}}^{\dagger T}\right)+H.c..
\end{equation}

Rotating to the basis that also diagonalizes Zeeman coupling term we get
\begin{align}
	\left[\mathcal{H}_{0}+\mathcal{H}_{Z}\right]_{k_{z}=0} & \approx\int d^{2}\mathbf{k}d_{\mathbf{k}}^{\dagger}\left[\left(E_{\mathbf{k}}-\mu\right)\nu_{0}-b^x_{\mathbf{k}}\nu_{z}\right]d_{\mathbf{k}}\nonumber \\
	\left[\mathcal{H}_{SC,1}\right]_{k_{z}=0} & \approx-\frac{\Delta_{1}}{2}\int d^{2}\mathbf{k}d_{\mathbf{k}}^{\dagger}\left(e^{-i\phi_{\mathbf{k}}}\nu_{x}\right)d_{-\mathbf{k}}^{\dagger T}+H.c..
\end{align}
The chemical potential will be crossing both conduction bands and leads to two split Fermi surfaces which have opposite relative helicity. The intra-orbital singlet pairing can only pair states on different Fermi surfaces. This suggests, as we confirm below, that the intra-orbital singlet pairing is suppressed by Zeeman couplings.

\subsection{Uniform orbital-triplet pairing}

The different types of pairings allowed by symmetries in regular centro-symmetric topological insulators
were previously studied in Ref.~\cite{PhysRevLett.105.097001}. In particular, a triplet-type pairing
deserves special attention here, which consists of inter-orbital orbital-triplet cooper pairs. In our basis choice, it reads as

\begin{align}
	\mathcal{H}_{SC,0} & =\frac{\Delta_{0}}{2}\int d^{2}\mathbf{k}\psi_{\mathbf{k}}^{\dagger}i\tau_{x}\sigma_{y}\psi_{-\mathbf{k}}^{\dagger T}+H.c.\nonumber \\
	& \approx\frac{\Delta_{0}}{2}\sum_{a,b=1,2}\int d^{2}\mathbf{k}f_{\mathbf{k},a}^{\dagger}\left(\varphi_{\mathbf{k}}^{a\dagger}i\tau_{x}\sigma_{y}\varphi_{-\mathbf{k}}^{b*}\right)f_{-\mathbf{k},b}^{\dagger}+H.c..
\end{align}

Using the unitary transformation $
\varphi_{\mathbf{k}}^{a\dagger}i\tau_{x}\sigma_{y}\varphi_{-\mathbf{k}}^{b*}=-e^{-i\phi_{\mathbf{k}}}\sin\alpha_{\mathbf{k}}\hat{e}_{a}^{T}\hat{e}_{b},
$ where $\sin\alpha_{\mathbf{k}}=\frac{\left|k\right|}{\sqrt{k^{2}+m_{k}^{2}}}$,
we get

\begin{equation}
	\mathcal{H}_{SC,0}=-\frac{\Delta_{0}}{2}\int d^{2}\mathbf{k}e^{-i\phi_{\mathbf{k}}}\sin\alpha_{\mathbf{k}}\left(f_{\mathbf{k}}^{\dagger}\nu_{0}f_{-\mathbf{k}}^{\dagger T}\right)+H.c..
\end{equation}
This pairing term commutes with the Zeeman term and so can be simultaneously diagonalized, returning

\begin{align}
	\left[\mathcal{H}_{0}+\mathcal{H}_{Z}\right]_{k_{z}=0} & \approx\int d^{2}\mathbf{k}d_{\mathbf{k}}^{\dagger}\left[\left(E_{\mathbf{k}}-\mu\right)\nu_{0}-b^x_{\mathbf{k}}\nu_{z}\right]d_{\mathbf{k}}\nonumber \\
	\left[\mathcal{H}_{SC,2}\right]_{k_{z}=0} & \approx-\frac{\Delta_{0}}{2}\int d^{2}\mathbf{k}e^{-i\phi_{\mathbf{k}}}\sin\alpha_{\mathbf{k}}\left(d_{\mathbf{k}}^{\dagger}\nu_{0}d_{-\mathbf{k}}^{\dagger T}\right)+H.c..
\end{align}

It is clear that, contrary to the intra-orbital singlet pairing, the inter-orbital triplet superconductivity pairs solely the states on each Zeeman-split helical Fermi surface, exclusively. As a result, the intra-orbital triplet pairing is not suppressed by the Zeeman coupling.

\subsection{Vortex Hamiltonian in inter-orbital triplet paired state}

Having in hands a type of pairing that is not suppressed by the Zeeman-induced
splitting of the Fermi surfaces, we can derive the
the spectrum of vortex bound states. We keep the axial symmetry of the problem and introduce a vortex along the $z$ direction, still focusing on $k_{z}=0$. We have
\begin{align}
	\left[\mathcal{H}_{SC,2}^{vrtx}\right]_{k_{z}=0} & =\frac{1}{2}\int d^{2}\mathbf{r}\psi^{\dagger}\Delta\left(\mathbf{r}\right)i\tau_{x}\sigma_{y}\psi^{\dagger T}+H.c.,
\end{align}
where
\begin{equation}
	\Delta\left(\mathbf{r}\right)=\frac{\Delta_{0}}{\xi_{0}}\left(x+iy\right).
\end{equation}

In momentum space and in polar coordinates

\begin{align}
	\left(\begin{array}{c}
		\partial_{k_{x}}\\
		\partial_{k_{y}}
	\end{array}\right) & =\left(\begin{array}{cc}
		\cos\phi_{\mathbf{k}} & -\sin\phi_{\mathbf{k}}\\
		\sin\phi_{\mathbf{k}} & \cos\phi_{\mathbf{k}}
	\end{array}\right)\left(\begin{array}{c}
		\partial_{k}\\
		\frac{1}{k}\partial_{\phi_{\mathbf{k}}}
	\end{array}\right).
\end{align}

Projecting the vortex Hamiltonian into the conduction bands we get
\begin{align}
	\left[\mathcal{H}_{SC,2}^{vrtx}\right]_{k_{z}=0} & =\frac{i}{2}\frac{\Delta_{0}}{\xi_{0}}\int d^{2}\mathbf{k}\psi_{\mathbf{k}}^{\dagger}\left(\partial_{k_{x}}+i\partial_{k_{y}}\right)i\tau_{x}\sigma_{y}\psi_{-\mathbf{k}}^{\dagger T}+H.c.\nonumber \\
	& \approx\frac{i}{2}\frac{\Delta_{0}}{\xi_{0}}\sum_{a,b=1,2}\int d^{2}\mathbf{k}f_{\mathbf{k},a}^{\dagger}\left[e^{i\phi_{\mathbf{k}}}\varphi_{\mathbf{k}}^{a\dagger}\left(\partial_{k}+i\frac{1}{k}\partial_{\phi_{\mathbf{k}}}\right)i\tau_{x}\sigma_{y}\varphi_{-\mathbf{k}}^{b*}\right]f_{-\mathbf{k},b}^{\dagger}+H.c..
\end{align}

The matrix elements associated with the Bloch wave-functions on each band read
\begin{align}
	& \varphi_{\mathbf{k}}^{a\dagger}\left(\partial_{k}+i\frac{1}{k}\partial_{\phi_{\mathbf{k}}}\right)i\tau_{x}\sigma_{y}\varphi_{-\mathbf{k}}^{b*}\\
	= & \left[i\left(\mathcal{A}_{k}\right)_{ab}+i\left(\mathcal{A}_{\phi_{\mathbf{k}}}\right)_{ab}\right]+\varphi_{\mathbf{k}}^{a\dagger}i\tau_{x}\sigma_{y}\varphi_{-\mathbf{k}}^{b*}\left(\partial_{k}+\frac{i}{k}\partial_{\phi_{\mathbf{k}}}\right)\nonumber 
\end{align}
where

\begin{align}
	i\left[\mathcal{A}_{k}\right]_{ab} & \equiv\varphi_{\mathbf{k}}^{a\dagger}i\tau_{x}\sigma_{y}\left(\partial_{k}\varphi_{-\mathbf{k}}^{b*}\right)\nonumber \\
	i\left[\mathcal{A}_{\phi_{\mathbf{k}}}\right]_{ab} & \equiv\varphi_{\mathbf{k}}^{a\dagger}i\tau_{x}\sigma_{y}\left(\frac{i}{k}\partial_{\phi_{\mathbf{k}}}\varphi_{-\mathbf{k}}^{b*}\right).
\end{align}

Since we solely consider states close to the Fermi surfaces in conduction bands $a,b=1,2$, the Berry connection-like terms read

\begin{align}
	i\left(\mathcal{A}_{k}\right)_{ab} & =-e^{-i\phi_{\mathbf{k}}}\frac{\partial_{k}\sin\alpha_{\mathbf{k}}}{2}\left(\hat{e}_{a}^{T}\hat{e}_{b}\right)\nonumber \\
	i\left(\mathcal{A}_{\phi_{\mathbf{k}}}\right)_{ab} & =-e^{-i\phi_{\mathbf{k}}}\frac{\sin\alpha_{\mathbf{k}}}{2k}\left(\hat{e}_{a}^{T}\hat{e}_{b}\right).
\end{align}

The vortex Hamiltonian then takes the form

\begin{align}
	\left[\mathcal{H}_{SC,2}^{vrtx}\right]_{k_{z}=0} & =-\frac{\Delta_{0}}{2\xi_{0}}\int d^{2}\mathbf{k}f_{\mathbf{k}}^{\dagger}i\nu_{0}\sin\alpha_{\mathbf{k}}\left[\partial_{k}+\mathcal{A}_{k}+\frac{i\partial_{\phi_{\mathbf{k}}}+\mathcal{A}_{\phi_{\mathbf{k}}}}{k}\right]f_{-\mathbf{k}}^{\dagger T}+H.c.
\end{align}
with

\begin{align}
	\mathcal{A}_{k} & =\partial_{k}\left(\log\sqrt{\sin\alpha_{\mathbf{k}}}\right)\nu_{0}\nonumber \\
	\mathcal{A}_{\phi_{\mathbf{k}}} & =\frac{\nu_{0}}{2}.
\end{align}
Notice that the radial component of the Berry connection is a pure
gauge and can be neglected. Furthermore, the angular part vanishes
by Fermi statistics. Going to the Zeeman diagonal basis is again trivial
and the Hamiltonian reduces to 

\begin{align}
	\left[\mathcal{H}_{SC,2}^{vrtx}\right]_{k_{z}=0} & =-\frac{\Delta_{0}}{2\xi_{0}}\int d^{2}\mathbf{k}d_{\mathbf{k}}^{\dagger}i\nu_{0}\sin\alpha_{\mathbf{k}}\left[\partial_{k}+\frac{i\partial_{\phi_{\mathbf{k}}}}{k}\right]d_{-\mathbf{k}}^{\dagger T}+H.c..
\end{align}

\subsection{Spectrum of vortex modes}

We are now ready to fully determine the spectrum of the Caroli-de
Gennes modes, as well as their corresponding wavefunctions. Introducing
Nambu spinors as

\begin{equation}
	\Psi_{\mathbf{k}}=\left(\begin{array}{c}
		d_{\mathbf{k}}\\
		d_{-\mathbf{k}}^{\dagger T}
	\end{array}\right),
\end{equation}
the Hamiltonian at $k_{z}=0$ reads
\begin{equation}
	\mathcal{H}_{TISC}^{k_{z}=0}=\frac{1}{2}\int_{\mathbf{k}}\Psi_{\mathbf{k}}^{\dagger}\left(\begin{array}{cc}
		\left(E_{\mathbf{k}}-\mu\right)\nu_{0}-h_{\mathbf{k}}\nu_{z} & -i\frac{\Delta_{0}}{\xi_{0}}\sin\alpha_{\mathbf{k}}\nu_{0}\left[\partial_{k}+\frac{i\partial_{\phi_{\mathbf{k}}}}{k}\right]\\
		-i\frac{\Delta_{0}}{\xi_{0}}\sin\alpha_{\mathbf{k}}\nu_{0}\left[\partial_{k}-\frac{i\partial_{\phi_{\mathbf{k}}}}{k}\right] & -\left(E_{\mathbf{k}}-\mu\right)\nu_{0}+h_{\mathbf{k}}\nu_{z}
	\end{array}\right)\Psi_{\mathbf{k}}.
\end{equation}

We can preform a mode expansion as
\begin{equation}
	\Psi_{\mathbf{k}}=\sum_{n}U_{n,k}\tilde{\Psi}_{n,k}
\end{equation}
where
\begin{align}
	U_{n,k} & =\frac{e^{-in\phi_{\mathbf{k}}}}{\sqrt{2\pi}}\left(\begin{array}{cccc}
		1\\
		& 1\\
		&  & i\\
		&  &  & i
	\end{array}\right).
\end{align}

The Hamiltonian then takes the form

\begin{equation}
	H=\frac{1}{2}\sum_{n,n'}\int\frac{dk}{2\pi}\tilde{\Psi}_{n,k}^{\dagger}\underbrace{\left[\int\frac{d\phi_{\mathbf{k}}}{2\pi}kU_{n,k}^{\dagger}h_{BdG}U_{n',k}\right]}_{\tilde{h}_{BdG}\delta_{nn'}}\tilde{\Psi}_{n',k},
\end{equation}
where
\begin{align}
	\tilde{h}_{BdG} & =\left(\begin{array}{cc}
		\left(E_{\mathbf{k}}-\mu\right)\nu_{0}-b^x_{\mathbf{k}}\nu_{z} & \frac{\Delta_{0}}{\xi_{0}}\sin\alpha_{\mathbf{k}}\nu_{0}\left(\partial_{k}+\frac{n}{k}\right)\\
		\frac{\Delta_{0}}{\xi_{0}}\sin\alpha_{\mathbf{k}}\nu_{0}\left(-\partial_{k}+\frac{n}{k}\right) & -\left(E_{\mathbf{k}}-\mu\right)\nu_{0}+b^x_{\mathbf{k}}\nu_{z}
	\end{array}\right).
\end{align}
We have two decoupled $\nu$ sectors: 
\begin{equation}
	\tilde{\Psi}_{n,k}=\sum_{\nu=\pm}\Theta_{n,k}^{\nu}a_{n,k}^{\nu},
\end{equation}
where 
\begin{equation}
	\Theta_{n,k}^{+}=\left(\begin{array}{c}
		u_{n}^{+}\left(k\right)\\
		0\\
		v_{n}^{+}\left(k\right)\\
		0
	\end{array}\right),\ \Theta_{n,k}^{-}=\left(\begin{array}{c}
		0\\
		u_{n}^{-}\left(k\right)\\
		0\\
		v_{n}^{-}\left(k\right)
	\end{array}\right)
\end{equation}
and
\begin{equation}
	\tilde{h}_{BdG}^{\nu}\Theta_{n,k}^{\nu}=E_{n}^{\nu}\Theta_{n,k}^{\nu}.
\end{equation}
The Nambu constraint enforces

\begin{equation}
	u_{n}^{\nu}\left(k\right)a_{n,k}^{\nu}=-i\left(-1\right)^{n}v_{-n}^{\nu*}\left(k\right)a_{-n,k}^{\nu\dagger}.
\end{equation}
Introducing $\Pi$ Pauli matrices in the Nambu space, the BdG Hamiltonian
for each set of modes reads
\begin{equation}
	\tilde{h}_{BdG}^{\nu}=\Pi_{z}\underbrace{\left(k-b^x_{\mathbf{k}}\nu-\mu\right)}_{\equiv\Delta_{0}\lambda_{\nu}\left(k\right)/\xi_{0}}+\Pi_{y}\frac{\Delta_{0}}{\xi_{0}}\left(i\partial_{k}\right)+\Pi_{x}\frac{\Delta_{0}}{\xi_{0}}\left(\frac{n}{k}\right)
\end{equation}
and we can solve for the lowest energy eigenstates by a Jackiw-Rebbi
argument for each $\nu$. The first two terms above have a zero-energy
mode when $k_{F}^{\nu}-h_{k_{F}^{\nu}}\nu=\mu$ so that we expect
to have the lowest energy states localized close to each Fermi surface.
For these modes
\begin{equation}\label{fne}
	E_{n}^{\nu}=\frac{\Delta_{0}}{\xi_{0}k_{c}^{\nu}}n,\,n\geq0.
\end{equation}
A zero-energy solution always exists for each $\nu$. The negative
$n$ modes are not independent from the positive $n$ (which compensates
the factor of $1/2$ in $H$).

Explicitly, the wavefunctions are fixed by
\begin{align}
	\left[\Pi_{z}\lambda_{\nu}\left(k\right)+\Pi_{y}\left(i\partial_{k}\right)\right]\Theta_{n,k}^{\nu} & =0\nonumber \\
	\Rightarrow\left[\partial_{k}+\Pi_{x}\lambda_{\nu}\left(k\right)\right]\Theta_{n,k}^{\nu} & =0,
\end{align}
with solutions
\begin{equation}
	\Theta_{n,k}^{\nu}\approx\frac{e^{-\int_{k_{F}^{\nu}-k}^{k_{F}^{\nu}+k}dk'\lambda_{\nu}\left(k'\right)}}{\sqrt{2\mathcal{N}}}\left(\begin{array}{c}
		1\\
		1
	\end{array}\right),
\end{equation}
where $\mathcal{N}$ is a normalization factor from the radial momentum
integration. In other words,
\begin{equation}
	u_{n}^{\nu}\left(k\right)=v_{n}^{\nu}\left(k\right)=\frac{e^{-\int_{k_{F}^{\nu}-k}^{k_{F}^{\nu}+k}dk'\lambda_{\nu}\left(k'\right)}}{\sqrt{2\mathcal{N}}}
\end{equation}
for the low energy modes. Also note the exponential localization of
the wavefunctions at $k_{F}^{\nu}$. Generically we will approximate
the solutions by evaluating them at the Fermi momenta. Notice that,
since the wavefunctions for the low energy modes are independent of
$n$ and real, the Nambu constraint reduces to
\begin{align}
	a_{-n,k}^{\nu} & =-i\left(-1\right)^{n}a_{n,k}^{\nu\dagger}\nonumber \\
	\Rightarrow a_{0,k}^{\nu} & =-ia_{0,k}^{\nu\dagger}.
\end{align}
Absorbing a $\pi/4$ phase in the operators, the $n=0$ modes corresponds to Majorana fermions.

We arrive at the expansion

\begin{align}
	d_{\mathbf{k}}^{\nu} & =\sum_{n}\frac{e^{-in\phi_{\mathbf{k}}}}{\sqrt{2\pi k}}u_{n}^{\nu}\left(k\right)a_{n,k}^{\nu}\approx f^{\nu}\left(k\right)\sum_{n}\left(e^{-in\phi_{\mathbf{k}}}a_{n,k}^{\nu}\right),
\end{align}
with 
\[
f^{\nu}\left(k\right)\equiv\frac{e^{-\int_{k_{F}^{\nu}-k}^{k_{F}^{\nu}+k}dk'\lambda_{\nu}\left(k'\right)}}{\sqrt{4\pi\mathcal{N}k}}.
\]

Now we may re-express the projected standard Fermion operators as
\begin{align}
	\psi_{\mathbf{k}} & \approx\sum_{a=1}^{2}\varphi_{\mathbf{k}}^{a}\left(e^{-i\nu_{y}\pi/4}d_{\mathbf{k}}\right)_{a}\nonumber \\
	& \approx\frac{1}{\sqrt{2}}\left[\varphi_{\mathbf{k}}^{1}\left(d_{\mathbf{k}}^{+}-d_{\mathbf{k}}^{-}\right)+\varphi_{\mathbf{k}}^{2}\left(d_{\mathbf{k}}^{+}+d_{\mathbf{k}}^{-}\right)\right].
\end{align}
Using $\varphi_{\mathbf{k}}^{a}$ at $k_{z}=0$ and the expansion
of the $d_{\mathbf{k}}^{\nu}$ operators, we obtain
\begin{align}
	\psi_{\mathbf{k}} & \approx\sum_{n}\left[\left(\begin{array}{c}
		0\\
		e^{-i\left(n-1\right)\phi_{\mathbf{k}}}\cos\frac{\alpha_{\mathbf{k}}}{2}\\
		e^{-in\phi_{\mathbf{k}}}\sin\frac{\alpha_{\mathbf{k}}}{2}\\
		0
	\end{array}\right)f^{+}\left(k\right)a_{n,k}^{+}-\left(\begin{array}{c}
		e^{-in\phi_{\mathbf{k}}}\cos\frac{\alpha_{\mathbf{k}}}{2}\\
		0\\
		0\\
		e^{-i\left(n-1\right)\phi_{\mathbf{k}}}\sin\frac{\alpha_{\mathbf{k}}}{2}
	\end{array}\right)f^{-}\left(k\right)a_{n,k}^{-}\right].
\end{align}

We finish by going to real space by Fourier transforming, $\psi\left(\mathbf{r}\right)=\int_{\mathbf{k}}e^{i\mathbf{r}\cdot\mathbf{k}}\psi_{\mathbf{k}}$.
The angular integrals introduce Bessel functions; using the definitions
of the $\alpha_{\mathbf{k}}$ angles and the exponential localization
of the wavefunctions at the Fermi surfaces, we finally arrive at

\begin{align*}
	\psi\left(\mathbf{r}\right) & \approx\sum_{\nu=1,2}\sum_{l}c_{l}^{\nu}\chi_{l}^{\nu}\left(\mathbf{r}\right)a_{l,k_{F}}^{\nu},
\end{align*}
where
\begin{align}\label{states}
	\chi_{l}^{+}\left(\mathbf{r}\right) & =\frac{1}{\sqrt{N_{k_{F}^{+}}^{+}}}\left(\begin{array}{c}
		0\\
		e^{-i\left(l-1\right)\theta}J_{l-1}\left(k_{F}^{+}r\right)\left(m_{k_{F}^{+}}+E_{k_{F}^{+}}\right)\\
		e^{-il\theta}J_{l}\left(k_{F}^{+}r\right)k_{F}^{+}\\
		0
	\end{array}\right)\nonumber \\
	\chi_{l}^{-}\left(\mathbf{r}\right) & =\frac{1}{\sqrt{N_{k_{F}^{-}}^{-}}}\left(\begin{array}{c}
		e^{-il\theta}J_{l}\left(k_{F}^{-}r\right)k_{F}^{-}\\
		0\\
		0\\
		e^{-i\left(l-1\right)\theta}J_{l-1}\left(k_{F}^{-}r\right)\left(m_{k_{F}^{-}}-E_{k_{F}^{-}}\right)
	\end{array}\right),
\end{align}
and $c_{l}^{\nu}=(-1)^lk^{\nu}_Ff^{\nu}(k^{\nu}_F)$.

\section{1D vortex Hamiltonian}

In order to explicitly examine the structure of vortex modes localized at the end of the vortices on the surface, we need to derive the effective vortex Hamiltonian which can be applied for $k_z\neq 0$. With the wavefunctions derived in last section, finite $k_{z}$ may be considered perturbatively. Considering the linear $k_{z}$ term in $H_{k_{z}}\approx-i\alpha_{z}\partial_{z}$,
we obtain
\begin{align}
	& \int d^{2}r\chi_{l'}^{-}\left(\mathbf{r}\right)\tau_{x}\sigma_{z}\chi_{l}^{+}\left(\mathbf{r}\right)\nonumber \\
	= & \delta_{l,l'}\int rdr\left[J_{l}\left(k_{F}^{+}r\right)J_{l}\left(k_{F}^{-}r\right)k_{F}^{+}k_{F}^{-}-J_{l-1}\left(k_{F}^{+}r\right)J_{l-1}\left(k_{F}^{-}r\right)\left(m_{k_{F}^{+}}+E_{k_{F}^{+}}\right)\left(m_{k_{F}^{-}}-E_{k_{F}^{-}}\right)\right]\nonumber \\
	\equiv & \tilde{\Delta}_{l}\neq0.
\end{align}
So,
\begin{equation}
	H_{k_{z}}\approx-i\tilde{\Delta}\eta_{x}\partial_{z}.
\end{equation}where $\mu_i$ are the Pauli matrices acting in the space of the two states given in equation \eqref{states}. As outlined in the main text, this linear gapless Hamiltonian does not support localized states at the ends of the vortex at zero energy.

To derive the vortex Hamiltonian for higher energy vortex modes, we have to consider the other matrix element associated with the term $\epsilon k_z^2 \beta$. The term is diagonal in the bases of states in equation \eqref{states} and its matrix elements are given by 

\begin{align}
	& \epsilon^+=\int d^{2}r\chi_{l'}^{+}\left(\mathbf{r}\right)\tau_{z}\sigma_{0}\chi_{l}^{+}\left(\mathbf{r}\right)\nonumber \\
	= &\delta_{l,l'}\int rdr\left[J_{l-1}^2(k_F^+ r)(m_{k_F^+}+E_{k_F^+})^2-J_l(k_F^+ r){k_F^+}^2\right]\\
	& \epsilon^-=\int d^{2}r\chi_{l'}^{-}\left(\mathbf{r}\right)\tau_{z}\sigma_{0}\chi_{l}^{-}\left(\mathbf{r}\right)\nonumber \\
	= &\delta_{l,l'}\int rdr\left[-J_{l-1}^2(k_F^- r)(m_{k_F^-}-E_{k_F^-})^2+J_l(k_F^- r){k_F^-}^2\right]
\end{align}
Note that $\epsilon^+>\epsilon^-$. In fact, for chemical potential close to regions where $m_{k_F}\approx 0$ (which is particularly relevant for experiments on FTS vortex structure) and for small Zeeman field, $\epsilon^+\approx-\epsilon^->0$. As a result, the second derivative term, leads to a a term of the approximate form
\begin{equation}
	-\bar{\epsilon}\eta_z \partial_z^2
\end{equation}where $\bar{\epsilon}$ has the same sign as $\epsilon$.

The the energy of vortex modes with finite energy add naturally to the 1D model as
$\frac{E_n^++E_n^-}{2}\eta_0+\frac{E_n^+-E_n^-}{2}\eta_z$. The stability of finite energy states at the end of vortex would then dependent on the sign of $E_n^+-E_n^-$. This can be readily checked from equation \eqref{fne}, which gives an inverse proportionality of finite energies with corresponding Fermi wave vector. The Fermi wave vector is given by $k_{F}^{\nu}=\mu+h_{k_{F}^{\nu}}\nu$ where $h_{{k_F}^\nu}=-\frac{\Delta_z}{2}\frac{m_{k_F^\nu}}{\mu}$. It is then readily clear that for $m_k>0$, $h_{k_F}^\mu<0$ so that $k_{F}^{+}<k_{F}^{-}$, leading to $E_n^+>E_n^-$. As is outlined in the main letter, this relationship leads to stability of finite energy states.

\section{Effect of Zeeman coupling on the type of superconducting state}

For a careful study of the energetics of the problem and which pairing
symmetry is favored in the presence of a Zeeman coupling, we perform
a mean-field self-consistent analysis. For this, we start with the
wavefunctions at finite $k_{z}$ and in the presence of Zeeman coupling.
This is slightly more involved.

Now we consider the full TI plus Zeeman Hamiltonian with all momenta
\begin{equation}
	\mathcal{H}=\mathcal{H}_{TI}+\mathcal{H}_{Z}.
\end{equation}
Using the usual squaring trick on $H\equiv H_{\mathbf{k}}+\frac{\Delta_Z}{2}\sigma_{z}$,
we may find the spectrum to be 
\begin{equation}
	\pm E_{\pm,\mathbf{k}}=\pm\sqrt{E_{\mathbf{k}}^{2}+\left(\frac{\Delta_Z}{2}\right)^{2}\pm\left|\Delta_Z\right|\sqrt{k_{z}^{2}+m_{\mathbf{k}}^{2}}}=\pm\sqrt{k_{\perp}^{2}+\left(\xi_{\mathbf{k}}\pm\frac{\Delta_Z}{2}\right)^{2}},
	\label{EnergyVsK}
\end{equation}
where we identified the TI energy spectrum $E_{\mathbf{k}}\equiv\sqrt{k^{2}+m_{\mathbf{k}}^{2}}$
but also defined $\xi_{\mathbf{k}}=\sqrt{k_{z}^{2}+m_{\mathbf{k}}^{2}}$
and rewrote the expression in a way that may suggest us some hints
on how to proceed. First let us rotate away the in-plane momenta,
\begin{equation}
	H'=e^{i\sigma_{z}\phi_{\mathbf{k}}/2}He^{-i\sigma_{z}\phi_{\mathbf{k}}/2}=\tau_{x}\sigma_{x}k_{\perp}+\tau_{x}\sigma_{z}k_{z}+m_{\mathbf{k}}\tau_{z}\sigma_{0}+\frac{\Delta_Z}{2}\tau_{0}\sigma_{z}.
\end{equation}
Then we rotate $k_{z}$ and mass terms,
\begin{align}
	\tau_{x}\sigma_{z}k_{z}+m_{\mathbf{k}}\tau_{z}\sigma_{0} & =\xi_{\mathbf{k}}\tau_{z}\sigma_{0}e^{i\tau_{y}\sigma_{z}\beta_{\mathbf{k}}}\nonumber \\
	\tan\beta_{\mathbf{k}} & =\frac{k_{z}}{m_{\mathbf{k}}}
\end{align}
So
\begin{equation}
	H''=e^{i\tau_{y}\sigma_{z}\beta_{\mathbf{k}}/2}H'e^{-i\tau_{y}\sigma_{z}\beta_{\mathbf{k}}}=\tau_{x}\sigma_{x}k_{\perp}+\xi_{\mathbf{k}}\tau_{z}\sigma_{0}+\frac{\Delta_Z}{2}\tau_{0}\sigma_{z}.
\end{equation}
Now since we rotated away $k_{z}$, $H''$ now commutes with the mirror
operation 
\begin{equation}
	\mathcal{M}=\tau_{z}\sigma_{z}.
\end{equation}
So we can project with
\begin{equation}
	P^{\nu}=\frac{1+\nu\mathcal{M}}{2},
\end{equation}
where $\nu=\pm$. We obtain
\begin{align*}
	H_{\nu}'' & \equiv P^{\nu}H''P^{\nu}\\
	& =\left(\frac{1+\nu\mathcal{M}}{2}\right)\left(\tau_{x}\sigma_{x}k_{\perp}+\xi_{\mathbf{k}}\tau_{z}\sigma_{0}+\frac{\Delta_Z}{2}\tau_{0}\sigma_{z}\right)\left(\frac{1+\nu\mathcal{M}}{2}\right)\\
	& =\left(\frac{\tau_{x}\sigma_{x}-\nu\tau_{y}\sigma_{y}}{2}\right)k_{\perp}+\left(\xi_{\mathbf{k}}+\nu\frac{\Delta_Z}{2}\right)\left(\frac{\tau_{z}\sigma_{0}+\nu\tau_{0}\sigma_{z}}{2}\right)\\
	& \equiv\Lambda_{x}^{\nu}k_{\perp}+\Lambda_{z}^{\nu}\left(\xi_{\mathbf{k}}+\nu\frac{h}{2}\right)
\end{align*}
where we introduced the $\Lambda^{\nu}$ $4\times4$ matrices according
to the terms in parenthesis; they satisfy an SU(2) algebra. Now the
matrices anti-commute and we can just finish as usual. We  write
\begin{align}
	H_{\nu}'' & =\Lambda_{z}^{\nu}\left[\Lambda_{z}^{\nu}\Lambda_{x}^{\nu}k_{\perp}+\left(\xi_{\mathbf{k}}+\nu\frac{\Delta_Z}{2}\right)\right]\nonumber \\
	& =E_{\nu,\mathbf{k}}\Lambda_{z}^{\nu}e^{i\Lambda_{y}^{\nu}\gamma_{\mathbf{k}}^{\nu}},
\end{align}
where 
\begin{equation}
	\tan\gamma_{\mathbf{k}}^{\nu}\equiv\frac{k_{\perp}}{\xi_{\mathbf{k}}+\nu\frac{h}{2}}
\end{equation}
and, by definition,
\begin{equation}
	\Lambda_{z}^{\nu}\Lambda_{x}^{\nu}=i\Lambda_{y}^{\nu}=\frac{i}{2}\left(\tau_{y}\sigma_{x}+\nu\tau_{x}\sigma_{y}\right).
\end{equation}

The wavefunctions now read
\begin{equation}
	\varphi_{\mathbf{k}}^{a_{\nu},\nu}=e^{-i\sigma_{z}\phi_{\mathbf{k}}/2}e^{-i\tau_{y}\sigma_{z}\beta_{\mathbf{k}}/2}P^{\nu}e^{-i\left(\frac{\tau_{y}\sigma_{x}+\nu\tau_{x}\sigma_{y}}{2}\right)\gamma_{\mathbf{k}}^{\nu}}e^{a_{\nu}}.
\end{equation}
Notice that the $a$ variable here is enslaved by $\nu$. For $\nu=+$,
$a_{+}=1,4$ and for $\nu=-$, $a_{-}=2,3$ which gives our four states
with energies $\left(-1\right)^{a_{\nu}+1}E_{\nu,\mathbf{k}}$.

Considering only the the positive energy bands we define the corresponding projection operators
\begin{align}
	\mathcal{P}^\nu_\mathbf{k}=&\frac{1}{4}\left(\tau_0\sigma_0+\nu\cos\beta_\mathbf{k}\tau_z\sigma_z+\nu\sin\beta_\mathbf{k}\tau_x\sigma_0+\cos\beta_\mathbf{k}\cos\gamma^\nu_\mathbf{k}\tau_z\sigma_0+\sin\beta_\mathbf{k}\cos\gamma^\nu_\mathbf{k}\tau_x\sigma_z \right. \nonumber \\
	&\left. + \nu \cos\gamma^\nu_\mathbf{k}\tau_0\sigma_z+\cos\phi_\mathbf{k}\sin\gamma^\nu_\mathbf{k}\tau_x\sigma_x+\sin\phi_\mathbf{k}\sin\gamma^\nu_\mathbf{k}\tau_x\sigma_y-\nu\cos\phi_\mathbf{k}\cos\beta_\mathbf{k}\sin\gamma^\nu_\mathbf{k}\tau_y\sigma_y \right. \nonumber \\
	&\left. +\nu\sin\phi_\mathbf{k}\cos\beta_\mathbf{k}\sin\gamma^\nu_\mathbf{k}\tau_y\sigma_x+\nu\cos\phi_\mathbf{k}\sin\beta_\mathbf{k}\sin\gamma^\nu_\mathbf{k}\tau_0\sigma_x+\nu\sin\phi_\mathbf{k}\sin\beta_\mathbf{k}\sin\gamma^\nu_\mathbf{k}\tau_0\sigma_y\right).
\end{align}
The single particle Green's function in the normal state for two positive bands can be written as
\begin{equation}
	G_0(i\omega_n,\mathbf{k})=\frac{\mathcal{P}^+_\mathbf{k}}{i\omega_n-\epsilon_{+,\mathbf{k}}}+\frac{\mathcal{P}^-_\mathbf{k}}{i\omega_n-\epsilon_{-,_\mathbf{k}}},
\end{equation}
where $\omega_n=(2n+1)\pi T$ are the Matsubara frequencies and $\epsilon_{\nu,\mathbf{k}}=E_{\nu,\mathbf{k}}-\mu$ ($\mu$ is the chemical potential). In order to write the linearized gap equation for the superconducting phase we define irreducible susceptibility as
\begin{equation}
	\chi_{ij}=-\frac{T}{N}\sum_{\omega_n,\mathbf{k}}\mathrm{Tr}\left[\frac{\hat{\Delta}_i}{\Delta}G_0(i\omega_n,\mathbf{k})\frac{\hat{\Delta}_j}{\Delta}G_0(-i\omega_n,\mathbf{k})\right],
	\label{susc}
\end{equation}
where $N$ is the number of unit cells, $\hat{\Delta}_i/\Delta$ defines the orbital and spin structure of different pairing potentials.
Following the work of Ref.~\cite{PhysRevLett.105.097001}, we consider only intra-orbital s-wave pairing and inter-orbital time-reversal invariant pairing channels. We treat the Zeeman field as a small perturbation compared to critical temperatures, so the phase diagram will still involve only these two phases, as was the case for $\Delta_Z=0$. For these, the orbital and spin structure have the following form
\begin{align}
	\hat{\Delta}_{1a}&=\Delta\tau_0\sigma_0,\qquad \mathrm{and} \qquad \hat{\Delta}_{1b}=\Delta\tau_z\sigma_0 \\
	\hat{\Delta}_0&=\Delta\tau_x\sigma_0.
\end{align}
Here, $\hat{\Delta}_{1a}$ and $\hat{\Delta}_{1b}$ are two different intra-orbital components for the s-wave channel and $\hat{\Delta}_0$ is the structure of inter-orbital pairing.

In order to evaluate the susceptibilities we make the following approximations: 1) that despite the presence of the Zeeman term the Fermi surface is approximately isotropic, 2) that the Zeeman field strength is small and expand the form of the energies of two Fermi surfaces (\ref{EnergyVsK})
\begin{equation}
	E_{\pm,\mathbf{k}}\approx E_\mathbf{k}\pm\frac{\Delta_Z\xi_k}{2E_\mathbf{k}},
	\label{EnergyExp}
\end{equation}
where $E_\mathbf{k}=\sqrt{k^2_\perp+\xi^2_k}$ is the energy of the bands when Zeeman term is zero. Now the integration with $\mathbf{k}$ of (\ref{susc}) can be carried for the case with Zeeman term. The only difference is that the Fermi energy where the integral is peaked will correspond to different momenta and different density of states for Zeeman split bands. Assuming that the density of states is the same for both bands and using the isotropy condition of Fermi surfaces, the Fermi momenta can be determined from the energy equation (\ref{EnergyVsK}). The only subtlety here arises for the cross term case in (\ref{susc}), which involves components from both bands. In that case we use the energy expansion (\ref{EnergyExp}) in the denominator and disregard the Zeeman term in the numerator. After that the sum for that case is similar to the case without Zeeman field, so the momenta where it is peaked is determined from $E_\mathbf{k}=\mu$. It should be noted that both the expansion of the energy (\ref{EnergyExp}) and the neglecting of Zeeman terms in the numerator of the cross term are valid for $\Delta_Z\ll\mu$ and $\Delta_Z\ll T$. After this, the susceptibilities related to $\Delta_1$ and $\Delta_0$ can be written as
\begin{align}
	\label{chi1}
	\chi_{2}&\equiv\chi_{1a1b}=\frac{\bar{\chi}}{2}\left[\frac{m^+_k}{\mu}\left(1+\frac{\Delta_Z}{2\xi^+_k}\right)+\frac{m^-_k}{\mu}\left(1-\frac{\Delta_Z}{2\xi^-_k}\right)\right]  \\
	\chi_{1}&\equiv\chi_{1b1b}=\frac{\bar{\chi}}{2}\left[\frac{(m^+_k)^2}{\mu^2}\left(1+\frac{\Delta_Z}{2\xi^+_k}\right)^2+\frac{(m^-_k)^2}{\mu^2}\left(1-\frac{\Delta_Z}{2\xi^-_k}\right)^2 +\left(\frac{k^0_\perp}{\xi^0_k}\right)^2\left(1-\frac{\mu^2-\frac{\Delta_Z^2}{4}}{\left(\mu+\frac{h\xi^0_k}{2\mu}\right)\left(\mu-\frac{\Delta_Z\xi^0_k}{2\mu}\right)}\right)\right] \\
	\chi_{0}&\equiv\chi_{00}=\frac{\bar{\chi}}{2}\left[\left(\frac{k^+_z}{\xi^+_k}\right)^2+\left(\frac{k^-_z}{\xi^-_k}\right)^2+\left(\frac{m^0_k}{\xi^0_k}\right)^2 \left(1-\frac{\left(\xi^0_k+\frac{\Delta_Z}{2}\right)\left(\xi^0_k-\frac{\Delta_Z}{2}\right)-(k^0_\perp)^2}{\left(\mu+\frac{\Delta_Z\xi^0_k}{2\mu}\right)\left(\mu-\frac{\Delta_Z\xi^0_k}{2\mu}\right)}\right)\right].
	\label{chi2}
\end{align}
The superscripts $+,-$ and $0$ denote that the corresponding quantities are evaluated at the energy $E_{+,\mathbf{k}}=\mu$, $E_{-,\mathbf{k}}=\mu$ and $E_{\mathbf{k}}=\mu$, respectively. $\bar{\chi}=-\int_{-w_D}^{w_D}\mathcal{D(\xi)}\tanh\left(\xi/2T\right)/2\xi d\xi$ is the standard $s$-wave susceptibility, $D(\xi)$ is the density of states and $w_D$ is the Debye frequency.
Plugging these forms of susceptibilities into linearized gap equations for s-wave and inter-orbital triplet pairings \cite{PhysRevLett.105.097001,nakosai2012,hashimoto2016}
\begin{equation}
	\label{LinGap}
	\mathrm{det}\left|\begin{array}{cc}
		U\bar{\chi}-1 & U\chi_{2} \\
		V\chi_{2} & V\chi_{1} -1
	\end{array}\right|=1, \qquad
	V\chi_{0}=1,
\end{equation}
we obtain the resulting phase boundary in this case (see Fig. 1 of the main text). In (\ref{LinGap}) $U$ and $V$ describe intra- and inter-orbital interactions of electrons, respectively \cite{PhysRevLett.105.097001,nakosai2012,hashimoto2016}.  

\putbib
\end{bibunit}
\end{document}